\documentclass[%
unsortedaddress,
preprint,
 amsmath,amssymb,
 aps,
]{revtex4-2}

\usepackage[utf8]{inputenc}
\usepackage[english]{babel}
\usepackage{titlesec}
\usepackage{etoolbox}
\usepackage{dcolumn}

\usepackage{amsmath}
\usepackage{amssymb} 

\usepackage{graphicx} 
\usepackage{xcolor}
\usepackage{natbib}
\usepackage{braket}

\usepackage{siunitx}

\usepackage{cancel}
\newcommand\Ccancel[2][black]{
    \let\OldcancelColor\CancelColor
    \renewcommand\CancelColor{\color{#1}}
    \cancel{#2}
    \renewcommand\CancelColor{\OldcancelColor}
}
\usepackage{bm}
\usepackage{float}
\renewcommand{\vec}{\bm}
\newcommand{\im}{\mathrm{i}}

\usepackage[normalem]{ulem}
\newcommand{\stkout}[1]{\ifmmode\text{\sout{\ensuremath{#1}}}\else\sout{#1}\fi}

\usepackage[all]{nowidow}

\begin{document}

\preprint{APS/123-QED}

\title{Electrically driven first-order phase transition of a 2D ionic crystal at the electrode/electrolyte interface}

\author{Federica Angiolari}
\affiliation{%
 Centre Européen de Calcul Atomique et Moléculaire (CECAM), Ecole Polytechnique Fédérale de Lausanne, 1015 Lausanne, Switzerland
}%
\affiliation{%
 National Centre for Computational Design and Discovery of Novel Materials (MARVEL), Ecole Polytechnique Fédérale de Lausanne,CH-1015 Lausanne, Switzerland
 }%

\author{Alessandro Coretti}%
\affiliation{%
Faculty of Physics, University of Vienna, 1090 Vienna, Austria
}%

\author{Mathieu Salanne}
\affiliation{
Sorbonne Universit\'e, CNRS, Physicochimie des Electrolytes et Nanosyst\'emes Interfaciaux, F-75005 Paris, France
}%
\affiliation{
Institut Universitaire de France (IUF), 75231 Paris, France
}%

\author{Sara Bonella}
 \email{sara.bonella@epfl.ch}
\affiliation{%
 Centre Européen de Calcul Atomique et Moléculaire (CECAM), Ecole Polytechnique Fédérale de Lausanne, 1015 Lausanne, Switzerland
}%
\affiliation{%
 National Centre for Computational Design and Discovery of Novel Materials (MARVEL), Ecole Polytechnique Fédérale de Lausanne,CH-1015 Lausanne, Switzerland
 }%

\date{\today}

\begin{abstract}
Liquid electrolytes adsorbed at the surface of metallic electrodes display a multitude of structures that can largely differ from the parent bulk system, both in terms of composition and local organization. In particular, 
the existence of disorder-order or order-order transitions has been increasingly reported in experimental and simulation studies, and the electrode potential identified as the corresponding driving force. The microscopic mechanisms and the stages of the process are, however, poorly understood, and the free energy variation during the transition remain insufficiently characterized. 
To fill this gap, we investigate the crystallization of the adsorbed layer in a prototypical molten salt-metal interface. We demonstrate that the transition from the disordered to ordered structures proceeds in two stages. Pre-ordering effects are observed across a wide range of potentials, resulting in the formation of a poly-crystalline structure on the electrode surface, before an abrupt ordering transition finally occurs. The pre-ordering displays signature of a continuous transition. On the other hand, finite-size effects analysis proves the first-order character of the transition towards the mono-crystalline state. Upon increasing the system size, a shift in the onset applied voltage is observed, accompanied by a dramatic increase in the free energy barrier. The latter reflects in the interfacial capacitance, which displays a peak that sharpens with increasing system size.
\begin{description}
\item[Significance Statement]
Electrochemical interfaces are key to technologies ranging from energy storage to catalysis, yet the fundamental mechanisms that govern their structural transformations remain poorly understood. In this work, molecular simulations reveal that applying an electrical potential can drive a two-stages transition from a disordered to an ordered interfacial state. The final transition is shown to be first-order in nature. These findings deepen our understanding of how electric fields influence matter at interfaces and open new possibilities for controlling nanoscale order in electrochemical systems.
\end{description}
\end{abstract}

\keywords{Order-disorder transition $|$ Electrochemical interface $|$ Molten salt $|$ Enhanced sampling}
\maketitle


\section{Introduction}
The formation of ordered anion adlayers on metal surfaces is a well-documented phenomenon that is known to affect the electrochemical reactivity at the interface.~\cite{magnussen2002a} Although original studies focused on halide anions, ordered structures were reported also for other systems, such as ionic liquids.~\cite{pan2006a,su2009a,wen2015a} Moreover, recent works showed the formation of ordered two-dimensional (2D) stoichiometric and non-stoichiometric crystals of NaCl on surfaces of graphene oxides,~\cite{shi2018a,yi2025a} as well as the existence of Wigner crystal-like structures of divalent chloride salts on the surface of polycrystalline gold, amorphous silica and gallium nitride.~\cite{wang2024a} These results suggest that ordering transitions of ionic species are very common at solid-liquid interfaces, and are not limited to the case of specific ions and/or perfect metallic surfaces.

A key aspect of these systems is the preeminence of Coulombic effects at the interface. The substrates are metallic, semi-metallic, or ionic, leading to strong interactions with the ions in the electrolyte. In addition, in the case of metals, the formation of ordered adlayers is often driven by the polarization of the electrode through the application of a voltage.~\cite{magnussen2002a} The role of electronic screening of the material in the interfacial crystallization properties of ionic liquids was also attested, showing that the stability of the ordered structure, measured via the mean solidification length of the liquid, is strongly enhanced by the use of highly polarizable surfaces~\cite{comtet2017a}. 

Despite these advances, critical gaps remain in understanding the mechanisms at play in these transitions. The above-mentioned studies generally combine surface-sensitive instruments such as scanning probe microscopes with measurements of macroscopic quantities such as the interfacial capacitance. These tools are very efficient for detecting transitions, but give little information on the structure of the adsorbed layer at the advent of ordering. In principle, theory and simulation should be effective tools for complementing experiments and providing additional microscopic details, but they have been hindered by the complexity of the systems considered. Some early studies characterized the ordering transition using a lattice-gas model,~\cite{wandlowski2001a} but so far very few molecular dynamics (MD) simulations reported the spontaneous formation of ordered structures on the surface of electrodes,~\cite{pounds2009,kirchner2013a,merlet2014} and their observations were hampered by limited statistics and the use of small simulation cells. Useful information on the stability of the adsorbed layers through the determination of accurate formation energies~\cite{shi2018a,yi2025a} can be achieved via electronic structure calculations, generally performed within the framework of density functional theory (DFT), but these calculations are too expensive for characterizing disordered states and free energy changes during the transitions.

In this study, we make a significant step towards the quantitative understanding of ionic ordering transition at metal surfaces, via classical constant-potential~\cite{siepmann_influence_1995, reed_electrochemical_2007} MD simulations of the electrochemical interface between polarizable molten LiCl and aluminum (Al) electrodes arranged in an FCC(100) structure. Although the simulated system cannot be considered a real system (aluminum would melt at the simulated temperature), Pounds et al.~\cite{pounds2009} showed that it exhibits an ordering structural transition upon application of a voltage and is therefore a good prototype for studying the nature of these transitions.
Previous simulations were limited to relatively small system sizes and short timescales that prevented full characterization of the adsorbed structures. They also  lacked analysis of finite-size effects. Indeed, small simulated systems can lead to significant bias in the mechanism and the kinetics of the phase transition. Here we perform simulation at three different sizes (labeled ``small'', ``medium'' and ``large'' in the following) and extend the timescale of the simulation by at least one order of magnitude, reaching trajectories of approximately 1~nanosecond when necessary. This was made possible by recent advances in the simulation of electrochemical interfaces, with new algorithms~\cite{coretti2020a,coretti2022a} and more efficient software~\cite{mwsoftware} that enable accurate and efficient handling of polarization effects in the electrode and the electrolyte at moderate computational costs. The atomistic nature of MD allows us to track the formation of ordered ionic structures at the interface over time, revealing the existence of preordering effects. The use of the electrode potential as a thermodynamic driving force, to which the local charge responds dynamically, enables calculating the free energy variations. The analysis of finite-size effects proves the first-order character of the transition.

\section{Results and Discussion}

\subsection{Interface order}
\begin{figure}
    \centering
\includegraphics[width=1\linewidth]{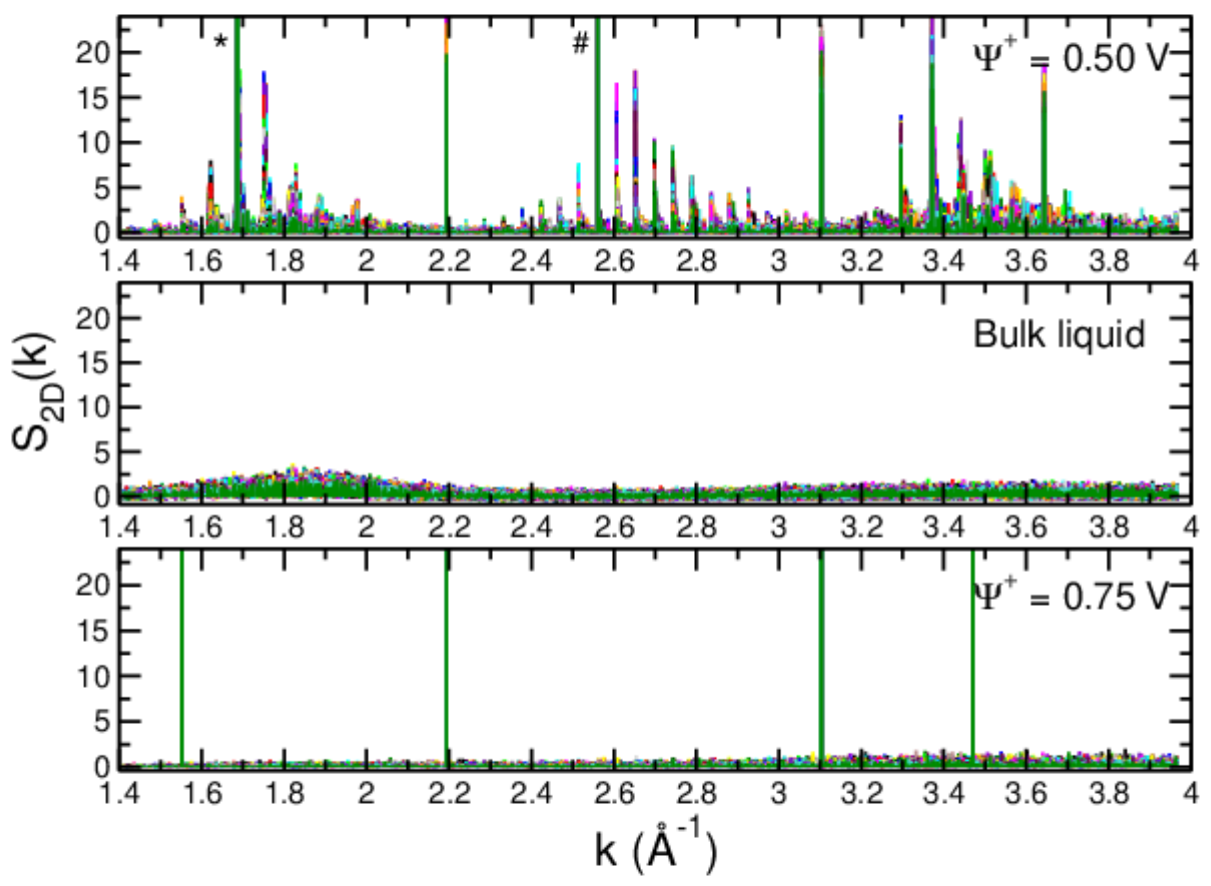}
\caption{Instantaneous in-plane structure factors for the adsorbed layer on the positive electrode set at 0.50~V (top) and 0.75~V (bottom), or in the bulk liquid (middle). Each color corresponds to a different configuration  over a trajectory of 300~ps, with a sampling time of 1~ps. For $\Psi^+$ = 0.50~V, the peaks labeled with $\star$ and $\#$ symbols have a larger intensity than the highest value shown on the plot, while this is the case for all the four peaks for $\Psi^+$ = 0.75~V (plots are zoomed in order to clearly show the multiple peaks at 0.50~V).}\label{fig:structurefactors}
\end{figure}

Unless stated otherwise, the results presented refer to the medium-sized simulation cell, which provides the best compromise between minimizing finite-size effects and maintaining computational efficiency. For each simulation, a potential difference $\Delta \Psi = \Psi^+ - \Psi^-$ is applied between the two electrodes. The two electrode potentials, $\Psi^+$ and $\Psi^-$, are known up to an arbitrary constant. Here, this constant is set so that $\Psi^+ = -\Psi^- = \Delta \Psi /2$. We begin by analyzing the spatial correlations of the electrolyte ions within the plane parallel to the electrode surface. We focus on the positive electrode, where pronounced ordering following the FCC lattice of aluminum is observed, consistent with the findings by Pounds et al.~\cite{pounds2009}. The ordering patterns observed in our simulations (larger system and longer time scales) are, however, considerably richer than those previously reported. To visualize the lateral ordering of the ions under different applied voltages, we monitor the 2D partial structure factor for Cl$^{-}$-Cl$^{-}$ pairs in the adsorbed layer:
\begin{equation}
S_{2D}(\vec{k})= \frac{1}{N_{\text{Cl}}}\sum_{i,j}\exp{\im \vec{k}\cdot \vec{r}_{ij}}
\end{equation}
\noindent where $N_{\text{Cl}}$ is the number of Cl$^-$ in the electrolyte layer adjacent to the electrode. (Similar behavior is observed for the $S_{2D}(\vec{k})$ of Li$^+$.) Very different results are obtained depending on the electrode potential. As shown in Figure \ref{fig:structurefactors}, for $\Psi^+$ below 0.75~V (top panel), the structure factor shows a multitude of Bragg-like peaks of intensities that vary along the simulation. For voltage at 0.75~V or above (bottom panel), only a few very distinct and intense Bragg peaks are observed throughout the whole trajectory. This is consistent with a disorder-order transition, albeit with the ``disordered'' state showing obvious signs of pre-ordering. For comparison, the 2D structure factors obtained in the bulk liquid (similar behavior for all voltages) are also shown in the middle panel. This is characterized by broad peaks that differ strongly from the adsorbed layer cases. Note that in the small-size simulation cell, and in agreement with the results in Ref.~\citenum{pounds2009}, the onset of ordering occurs at a lower applied voltage ($\Psi^+=0.4$~V), reflecting finite-size effects that will be further discussed in the following.

\begin{figure*}[htbp]
    \centering
    \includegraphics[width=1.0\textwidth]{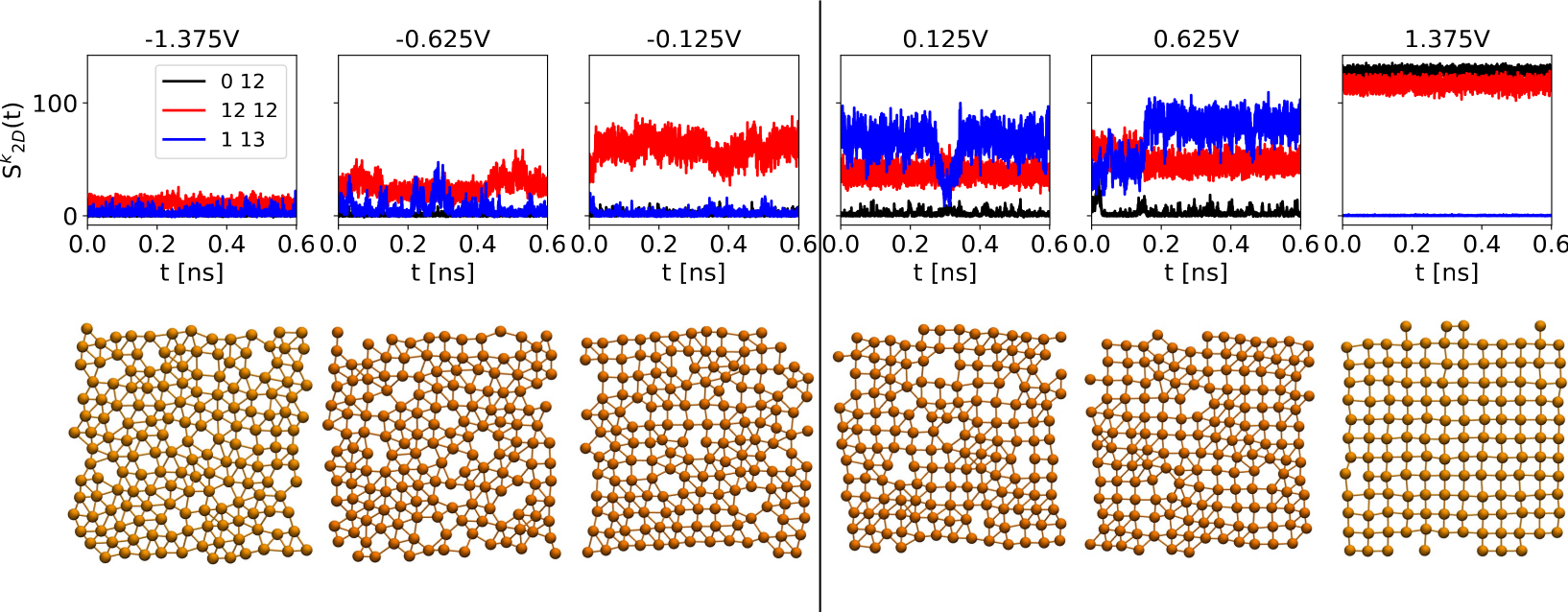}
    \caption{Top: Variation with time of the intensities of three characteristic 2D structure factor peaks ((0,12),(12,12) and (1,13)) for three negative (left) and positive (right) electrode potentials. Bottom: Representative instantaneous snapshots of the chloride ions positions at the same electrode potentials.}
    \label{fig:sk_snapshot}
\end{figure*}
In the 2D partial structure factor, we have identified three peaks that track the structural variations at the interface and reflect the extent of ordering quite precisely. These peaks can be distinguished via the vector $\vec{k}=\frac{2\pi}{L}(k^x,k^y)$ where $L$ is the lateral length of the simulation cell. They correspond to $(k^x,k^y)$ equal to (1,13), (0,12), and (12,12). Figure~\ref{fig:sk_snapshot} shows the temporal evolution of the intensity of the three characteristic peaks at both negative (left panels) and positive (right panel) electrode potentials, along with representative snapshots of Cl$^-$ configurations -- the same behavior is observed for Li$^+$ ions (Supplementary Information, Figure 2). At large negative potential (-1.375~V), the peak intensity is low for all the $\vec{k}$ vectors, and the adsorbed layer exhibits a highly disordered structure. We then observe a progressive increase of the intensity of the (12,12) and (1,13) peaks when the electrode potential grows up to $\Psi^+=0.625$~V. This indicates a partial preordering of the system, which is illustrated by the formation of well-organized domains and visible defects on the snapshots. Finally, for positive potentials above 0.625~V, the intensity of the (1,13) peak goes down to zero while the ones of the (0,12) and (12,12) peaks become maximal. A clear structural ordering is observed over a square lattice, as shown in the corresponding snapshot.

\begin{figure*}[hbp]  
    \centering
\includegraphics[width=\linewidth]{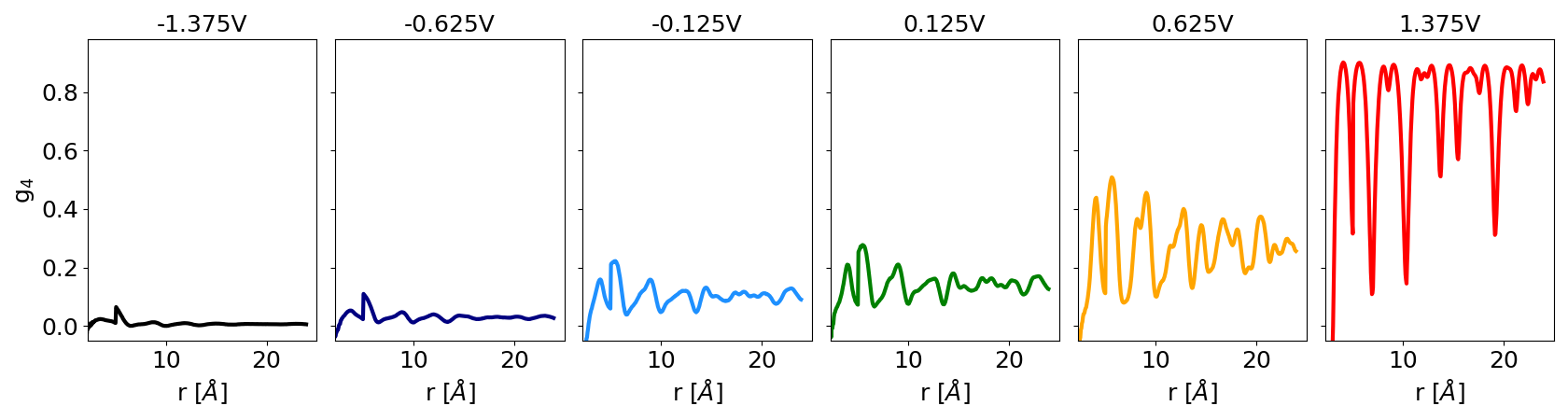}
\includegraphics[width=\linewidth]{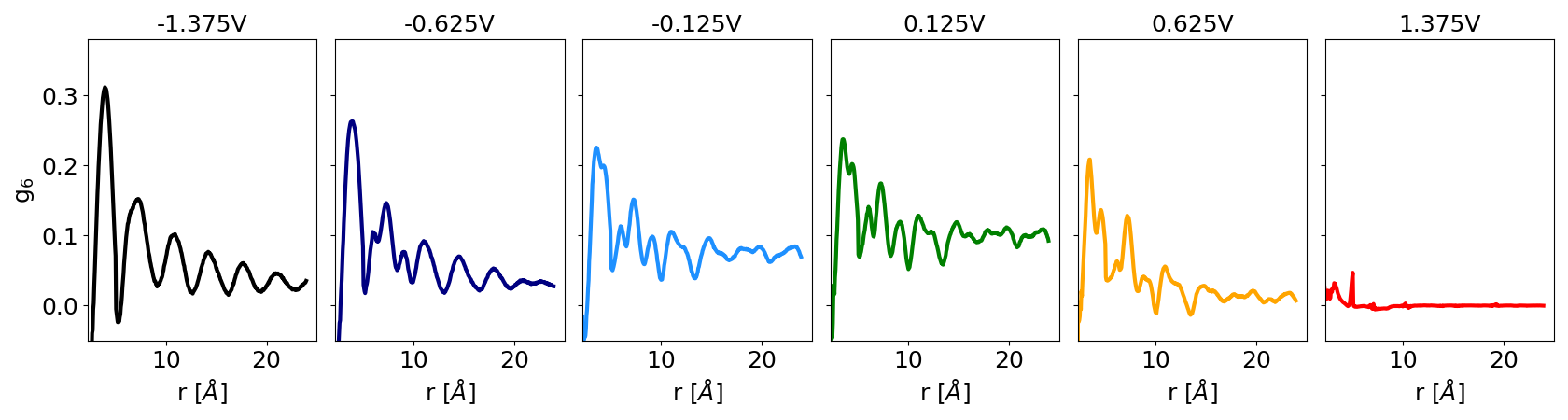}
    \caption{Orientational correlation functions of the adsorbed chloride ions, $g_4(r)$ (top) and $g_6(r)$ (bottom), computed for the different electrode potentials.}
    \label{fig:orientat_simmetry}
\end{figure*}

This phenomenology is reminiscent of the phase behavior of 2D crystals. For these systems, the Kosterlitz–Thouless–Halperin–Nelson–Young (KTHNY) theory predicts a two-step melting process occurring through the formation of an intermediate hexatic phase.~\cite{kosterlitz1973a,halperin1978a,young1979a} The two transitions are continuous, i.e. they occur smoothly without latent heat, and with no discontinuity in the first derivative of the free energy. However, many systems display deviations from the KTHNY theory, and the study of 2D crystals remains an intense field of study.~\cite{clark2009a,li2023b,loffler2015a} Borrowing from it, we compute the local orientational order parameter $\psi_n(i)$, for the $i$-th particle:
\begin{equation}
    \psi_n(i) = \frac{1}{N_i}\sum_{j=1}^{N_{i}}e^{in\theta_{ij}}
\end{equation}
In the equation above, $N_i$ are the nearest neighbors of the particle $i$ within a cut-off set to 5~\AA\ (i.e. the location of the first minimum of the Cl-Cl in-plane radial distribution function), $\theta_{ij}$ is the angle between particles $i$ and $j$ and, in our case, $n$ is $4$ or $6$ to investigate fourfold  or  sixfold angular symmetry, respectively. We then determined the  orientational correlation function 
\begin{equation}
    g_n(r) = \Braket{\psi_n(i)\psi_n^*(l)}.
\end{equation}
\noindent where $r$ is the distance between particles $i$ and $l$. These functions further characterize in-plane ordering by providing information about local angular correlation of nearby ions: if $g_n(r)$ is close to 1, the local neighborhoods of particles at a distance $r$ are oriented in the same direction, while if $g_n(r)$ is close to 0 the angular orientations are uncorrelated.  Figure~\ref{fig:orientat_simmetry} shows $g_4(r)$ and $g_6(r)$ for the chloride ions adlayer at the positive electrode for different potentials.

Three different regimes are observed. For positive electrode potentials above 0.75~V, $g_4$ remains close to 1 for the whole distance range (except for the drops at distances where there are no neighbours), while $g_6$ is null. This is consistent with the square geometry of the ordered 2D crystal phase. On the other hand, for potentials below $-0.5$~V, the $g_4$ function remains very close to 0, while the $g_6$ function shows a larger intensity ($\approx$~0.3) at low range ($r$~$\approx$~4~\AA\, corresponding to nearest-neighbour distances), followed by a decaying layer. In previous studies of 2D crystals, such a behavior was attributed to either the disordered liquid or to the hexatic phase. In principle, these can be distinguished by analyzing whether the large-$r$ decay follows an exponential or a power-law rule, but our simulated system is too small to conclude on this point, so in the following we will refer to this phase as being ``liquid-like''. Finally, for electrode potentials between $-0.5$~V and 0.75~V, the $g_4$ and $g_6$ functions both adopt finite values with no clear signs of decays over the studied range of distances. This shows that the structure of the adsorbed layer in this range of potentials consists in a mosaic of crystals with either tetratic or hexatic symmetry. The intensity of the functions varies progressively: Close to the liquid-like phase, both $g_4$ and $g_6$ take values close to 0.1, but as the potential approaches 0.75~V $g_4$ increases while $g_6$ starts to vanish, indicating a predominance of the tetratic-ordered nanocrystals at the onset of crystallization. Our observations indicate that the phase diagram of these 2D ionic crystals is much richer than expected from the KTHNY theory, most likely due to the importance of Coulombic interaction. 

Similar behavior is observed for the lithium cations sublattice at the positive electrode. Interestingly, no order is induced for either ion at the negative electrode. This asymmetric behavior can be attributed to the different interactions between the two ions and the surface. Chloride anions have a larger ionic radius and a large polarizability. As shown in previous work,~\cite{tazi} upon adsorption they display a strong dipole moment oriented towards the surface, with an intensity that increases with the applied potential. In contrast lithium ions are weakly polarizable so their interaction with the surface is less intense, which results in the persistence of a liquid-like structure at the negative electrode at all voltages. It could, however, be speculated that more symmetric ionic compounds such as NaCl or KCl would display a different behavior.

\subsection{Free Energy and capacitance}

\begin{figure}
\centering
\includegraphics[width=0.8\linewidth]{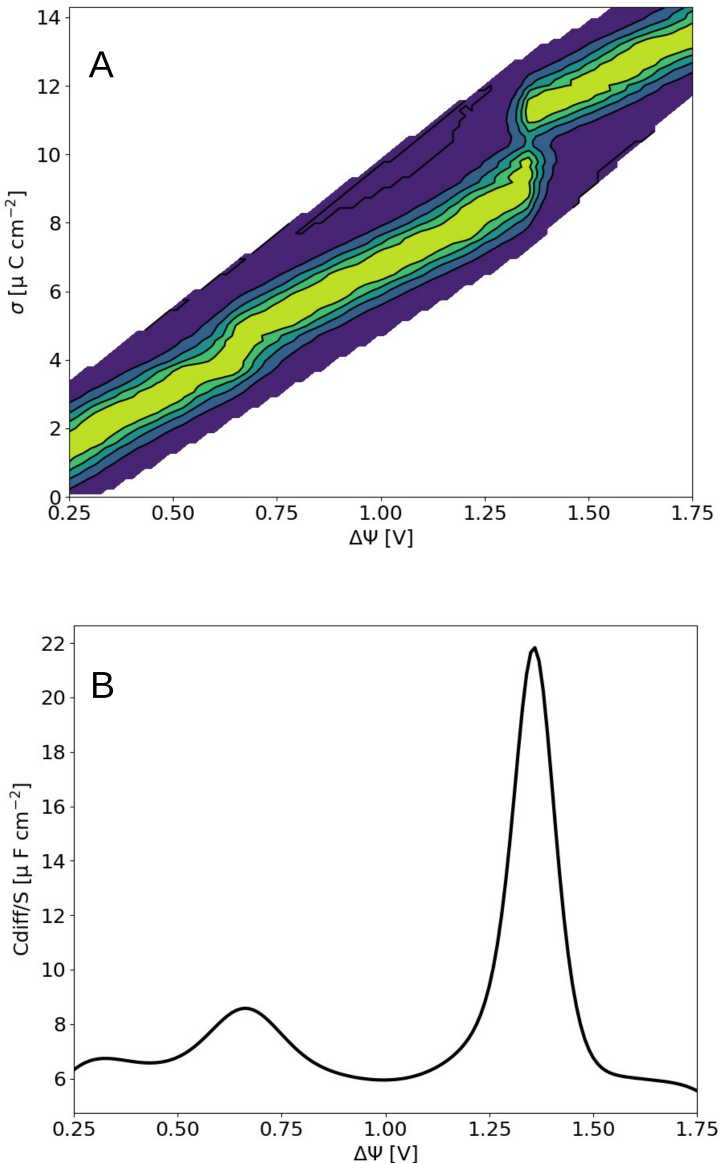}
\caption{(A) Probability $P(\sigma \mid\Delta\Psi)$ of the charge density on the electrodes at a given applied potential. (B) Differential capacitance as a function of applied voltage.  }\label{fig:fes}
\end{figure}
Two different scenarios are conceivable for the nature of the transitions: First-order (involving a sharp, discontinuous change with latent heat and phase coexistence) or crossover (involving a smooth, continuous change without singularities or latent heat). To discriminate, we study the behavior of the free energy of the system with respect to the thermodynamic driving force (i.e. the applied potential). As the system evolves under an applied potential, the surface charge density $\sigma$ reflects the reorganization of interfacial ions and can serve as an effective order parameter. We therefore calculated the probability distribution of surface charges at a given applied potential using the histogram reweighting technique.~\cite{limmer2013a} The results are presented in Figure~\ref{fig:fes}~A, where we observe two deviations from a linear behavior, for applied potentials of 0.7~V (labeled $\Delta \Psi_{\rm T1}$) and 1.35~V ($\Delta \Psi_{\rm T2}$). These are in good agreement with our observed structural changes in the equilibrium simulations: For $\Delta \Psi_{\rm T1}$, the negative electrode is at a potential of $-0.35$~V, close to the value for which we observed the liquid-like to polycrystal transition, while $\Delta \Psi_{\rm T2}$ corresponds to a positive electrode potential of 0.675~V, i.e. where the polycrystal to monocrystal transition occurs. These values are more accurate estimates of the transition voltage compared to the ones discussed in previous sections ($-0.5$~V and 0.75~V) because the enhanced sampling approach allows estimating the probability precisely for any applied potential.

We first analyze the latent energies associated with each transition. The surfacic energy associated with the capacitor is given by $\sigma\Delta\Psi$, so that the surfacic energy jump due to the polycrystal to monocrystal transition is 
\begin{equation}
\Delta E_{\rm T2}= \left(\sigma(\Delta \Psi_{\rm T2}^+)-\sigma(\Delta \Psi_{\rm T2}^-)\right) \Delta \Psi_{\rm T2}
\end{equation}
\noindent where 
\begin{eqnarray}
\sigma(\Delta \Psi_{\rm T2}^+)&=& \lim_{\Delta\Psi\rightarrow\Delta \Psi_{\rm T2}^+}\sigma(\Delta\Psi) \\
\sigma(\Delta \Psi_{\rm T2}^-)&=& \lim_{\Delta\Psi\rightarrow\Delta \Psi_{\rm T2}^-}\sigma(\Delta\Psi)
\end{eqnarray}
We obtain an estimate of $\Delta E_{\rm T2}=0.038$~J~m$^{-2}$. Normalizing this quantity by the surface concentration of ions $\gamma=10.1$~$\mu$mol~m$^{-2}$ (obtained by integrating the first peak of the density profiles at the transition potential) yields the latent energy of the transition, $ L=\Delta E_{\rm T2}/\gamma=3.4$~kJ~mol$^{-1}$. This value is approximately 6 times smaller than the one measured for the melting of the bulk 3D LiCl crystal (19.9~kJ~mol$^{-1}$)~\cite{lide2005crc}, which is consistent with the 2D geometry that reduces the cohesive energy and points towards a first-order character for this transition.

A second approach for characterizing the transitions is to evaluate the response function, which in the present case is the differential capacitance $C_{\text{diff}}$ of the interface.  The electrolyte contribution is given, as a function of the applied voltage, by:~\cite{limmer2013a}
\begin{equation}
    C_{\text{diff}} = \frac{\partial \langle \sigma \rangle}{\partial \Delta \Psi} = \frac{S}{k_B T} \langle (\delta \sigma)^2 \rangle
    \label{eq:capa}
\end{equation}
where $\delta \sigma = \sigma - \langle \sigma \rangle$, $S$ is the surface area, $k_B$ the Boltzmann constant and $T$ the temperature. Note that a contribution from the empty capacitor should be added to obtain absolute values,~\cite{scalfi2020a} but this term does not vary with system size (when normalized by the surface area), so it does not affect our analysis. Figure~\ref{fig:fes}~B shows that the capacitance variation is characterized by two anomalous peaks located at $\Delta \Psi_{\rm T1}$ and $\Delta \Psi_{\rm T2}$. 
The first peak has a relatively low intensity, while the second one is much sharper. Data for the small and large systems (see also next section) indicate that the second peak shifts to higher voltages, sharpens, and grows rapidly in intensity with increasing system size. Together with the data on the latent heat and the information gathered from the orientational order parameters, this indicates that the transition from liquid-like to polycrystal states is continuous (or weakly first-order), while the polycrystal to monocrystal transition can be classified as first-order.

\begin{figure}
\centering
\includegraphics[width=.8\linewidth]{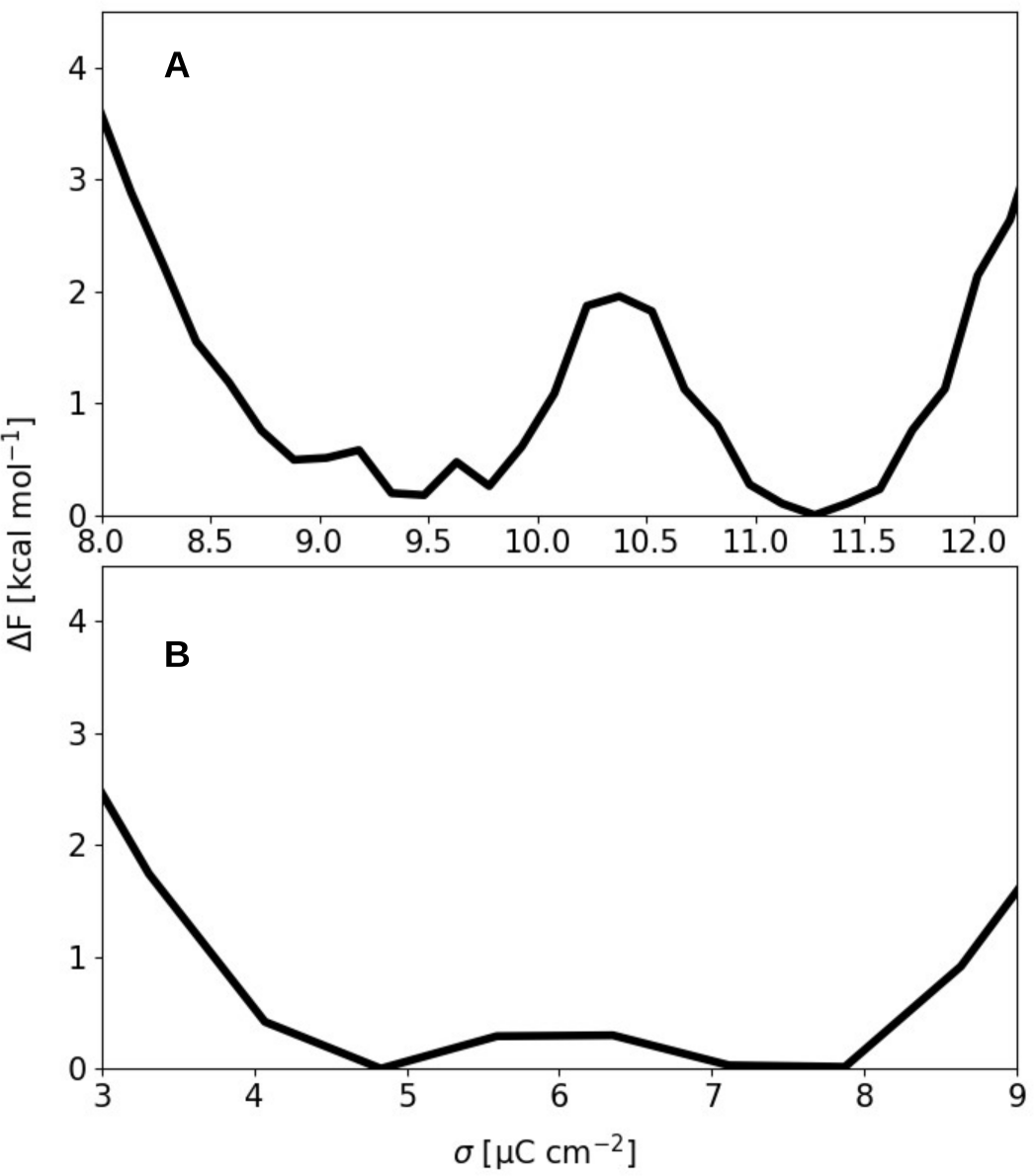}
\caption{Free energy profile for the medium (A) and small (B) size systems computed at the transition potential ($\Delta\Psi = 1.35$~V for the medium and $\Delta\Psi = 0.80$~V for the small one). }
\label{fig:finitesize}
\end{figure}

\subsection{Finite-size effects}  Transitions often occur via a nucleation process that involves the formation of an interface and that can be strongly affected by the finite-size of the system.~\cite{binder1987a} We therefore compare the results obtained with the small and large systems to the case of the medium system that was detailed so far. From the structural point of view, in the small system, the ordering transition is abrupt and observed at a much lower potential ($0.80$~V at the positive electrode), which is consistent with previous work by Pounds {\it et al.}~\cite{pounds2009} At lower electrode potentials, the 2D structure factors is reminiscent to the one of a disordered structure (Supplementary Figure 5). However, unlike the medium system, it is not possible to observe as explicitly the formation of a polycrystal. This is certainly indicative of finite-size effects as the surface is too small to accommodate several patches of coexisting crystals. Consistently, the surface charge probability shows a single deviation from linearity at applied potential 0.8~V (Supplementary Figure 5), accompanied by a single relatively broad peak for the differential capacitance. 

To compare the simulations at the two system sizes more quantitatively, we also consider how the free energy 
\begin{equation}
F^{\Delta \Psi}(\sigma) = -k_B T \ln P(\sigma \mid \Delta \Psi)
\end{equation}
\noindent 
($k_B$ is Boltzmann's constant and $T$ the temperature) changes with size at the transitions. Figure \ref{fig:finitesize} shows the free energy for the applied potential for which the two states are equally probable ($\Delta \Psi=0.8$~V for the small system and $\Delta \Psi = 1.35$~V for the medium system). In the small system, the free energy barrier separating the two states is very low (approximately 0.25~kcal/mol), making it difficult to resolve distinct minima. In contrast, in the medium-size system, the free energy surface reveals two well-defined minima separated by a free energy barrier of approximately 2~kcal/mol.
 This increase  with system size is consistent with a first-order transition. The barrier corresponds to the cost of creating an interface between coexisting phases, and this cost scales with the surface area of the interface, which grows with system size. Scaling further the size, for the large system we could not observe the transition, via brute force MD in the accessible computational time, between ordered and disordered structures: The system remained locked in one of the two phases, depending on the initial state. These extremely long lived metastabilities point to a further increase of the free energy barrier, consistent with the first-order character of the transition.

Analysis of the finite-size effects on the capacitance also points to a first-order transition at the positive electrode. Indeed, at coexistence of two phases, the capacitance is expected to increase with system size due to cooperative effects which span the entire surface. Our results show a peak in the capacitance at the transition potential for both the small {(see SI, Figure 6)} and medium system. For the medium system, however, the peak is more intense and sharper. Although we cannot identify the transition voltage for the large system due to the hysteresis effect mentioned above, we can estimate the capacitance for this system size by assuming the transition to occur at the same voltage as for the medium one (1.35~V) and simulating both the ordered and disordered states in independent runs. Equation~\ref{eq:capa} then yields $C_{\text{diff}} = 88.6$~$\mu$F~cm$^{-2}$, which provides a further confirmation of the growth of differential capacitance with system size.

\section{Conclusion}
While the existence of a voltage-driven first-order transition in molten LiCl at aluminum electrodes has been previously reported, our study offers a significantly deeper understanding of its microscopic and thermodynamic behavior. By computing and comparing free energy surfaces for different electrode sizes, we show how the system size affects the free energy landscape.  Our approach enables a much more accurate description of the transition, that occurs via two distinct stages instead of a single one as was previously reported. The simulations give direct access not only to the free energy barrier but also to the different ordering patterns at the surface, features hardly accessible in most experimental settings. 
By characterizing orientational ordering at the surface it is possible to observe complex structural motifs that evolve with the applied voltage, and could impact transport, reactivity and material selectivity in high-temperature electrochemical technologies. These results demonstrate that increasing system size enables a more accurate characterization of interfacial phase transitions, capturing stronger correlations and refining the determination of critical voltages. This has direct implications for technological applications, where molten salts are used in metal smelting, electrochemical recycling and next-generation molten salt reactors. For example, in electrochemical metal smelting, controlling the ordering of ions at the interface could enhance the selectivity of metal deposition, reducing energy consumption and impurity incorporation. Understanding these transitions at the molecular scale is thus critical for optimizing performance and extending the applicability of molten salt electrochemistry. From an energy storage point of view, the latent heat associated to the transition is relatively low. Alongside with the planar shape on the interface, this prevents the use of such systems for micro-supercapacitor devices. However, the charge jump we observe at the transition is significant, and it could enable the use of such interfaces in sensors or for reversibly triggering the motion of an actuator.

\section{Materials and Methods}
\subsection{MD simulation}
We perform three sets of classical molecular dynamics simulations of an Al-electrode/LiCl melt system, using the software MetalWalls~\cite{mwsoftware}. In the system referred to as ``small'' in the paper, the electrolyte consists of 500 Li and 500 Cl ions, while the electrodes are made of 432 fixed Al atoms arranged in an FCC structure. The size of the box is 24.25~\AA~ in the x and y directions, 57.93~\AA~ along the z-axis, with Periodic Boundary Conditions (PBC) applied in the x and y directions. To account for finite-size effects, we also study the same system with four times the surface area (``medium'' system in the text), for a total of 2000 Li and 2000 Cl ions, with 864 Al atoms for each electrode. Finally, an even larger system is considered, consisting of 8000 Li and 8000 Cl ions, with 3456 Al atoms per electrode, and a box size of 97~\AA~in the x and y directions, maintaining the same z-dimension and periodic boundary conditions (``large'' system in the text). NVT dynamics is performed using a Langevin thermostat with a friction coefficient of 0.1~ps$^{-1}$ and a timestep of 1.0~fs at 1200~K. The system is equilibrated for 100-175~ps, followed by 200-600~ps of dynamics.
 All simulations are performed in a constant potential ensemble~\cite{siepmann_influence_1995, reed_electrochemical_2007} and the explored applied voltage spans from 0.25~V to 2.75~V for the small and medium systems, the large system was studied at a single potential of 1.375~V. To increase the sampling of the phase space for the medium systems, at each applied voltage, two simulations are performed, starting from a configuration obtained from 100~ps at 2000~K.
The force field used in our simulation includes the polarization of the electrolyte~\cite{pounds2009}, ensuring the correct representation of the electrostatic properties. The charges on the electrodes are modeled via the fluctuating charge model~\cite{sprik,siepmann_influence_1995}: to each Al atom is assigned a Gaussian charge distribution, whose magnitude responds dynamically to the configuration of the electrolyte to keep a constant potential at the electrode atoms, thus sampling the constant potential ensemble. 
The interaction energy between melt ions and electrode atoms is described using the force field introduced in Refs.~\cite{pounds2009,tazi}, in which the parameters were obtained from DFT calculations using a generalized force-fitting strategy.
The dynamics is propagated using Mass Zero constraint dynamics (MaZe)~\cite{Ryckaert:1981,mazecom}, treating both the induced dipole moments on the melt ions and the electrode charges as additional variables, and enforcing global electroneutrality.

\subsection{Free energy} 
Free energy surfaces and barriers are computed using the Weighted Histogram Analysis Method (WHAM)~\cite{wham}. Following Ref.~\cite{merlet2014}, in this analysis the applied potential serves as the biasing variable and is maintained constant during each simulation.

\bibliographystyle{unsrt}
\bibliography{reference}

\clearpage
\onecolumngrid
\normalsize
\patchcmd{\large}{15}{15}{}{}
\begin{center}
  \textbf{\LARGE Supporting Information for\\ \textbf{Electrically driven first-order phase transition of a
2D ionic crystal at the electrode/electrolyte
interface}}\\[.2cm]
  Federica Angiolari,$^{1,2}$ Alessandro Coretti,$^{3}$ Mathieu Salanne$^{4,5}$ and Sara Bonella$^{1,2}$\\[.1cm]
  {\itshape
  ${}^1$Centre Européen de Calcul Atomique et Moléculaire (CECAM), Ecole Polytechnique Fédérale de Lausanne, 1015 Lausanne, Switzerland \\
  ${}^2$National Centre for Computational Design and Discovery of Novel Materials (MARVEL), Ecole Polytechnique Fédérale de Lausanne,CH-1015 Lausanne, Switzerland \\
  ${}^3$Faculty of Physics, University of Vienna, 1090 Vienna, Austria\\
  ${}^4$Sorbonne Universit\'e, CNRS, Physicochimie des Electrolytes et Nanosyst\'emes Interfaciaux, F-75005 Paris, France
  ${}^5$Institut Universitaire de France (IUF), 75231 Paris, France
  }
  ${}^*$Electronic address: sara.bonella@epfl.ch\\
(Dated: \today)\\[2cm]
\end{center}

\setcounter{equation}{0}
\setcounter{figure}{0}
\setcounter{table}{0}
\setcounter{page}{1}
\setcounter{section}{0}
\renewcommand{\theequation}{S\arabic{equation}}
\renewcommand{\thefigure}{S\arabic{figure}}
\renewcommand{\thetable}{S\arabic{table}}
\renewcommand{\bibnumfmt}[1]{[S#1]}
\renewcommand{\citenumfont}[1]{S#1}
\renewcommand{\thesection}{S\Roman{section}}
\renewcommand{\thepage}{S\arabic{page}}

\titleformat*{\section}{\Large\bfseries}

\section{Density plot}
\begin{figure}[H]
    \centering\includegraphics[width=0.7\textwidth]{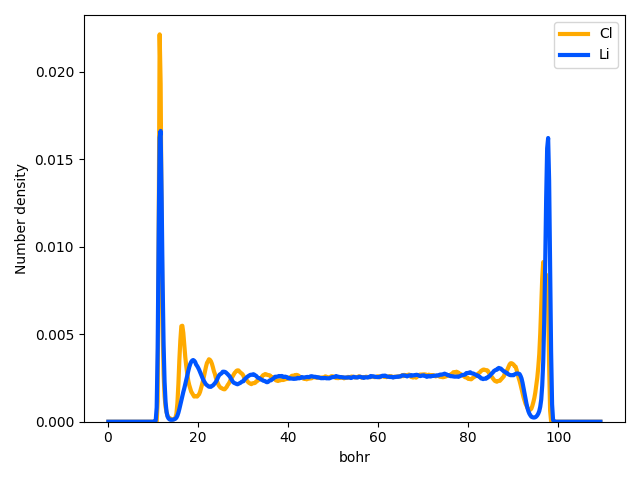}
    \caption{Density profiles of chloride and lithium across the simulation cell under an applied voltage of 1.5 V for the medium size system. These profiles, averaged over 75 ps, were used to calculate the latent heat by integrating the first interfacial peak.}
    \label{fig:neg_CLCL_big}
\end{figure}

\section{Structure Factor}

\begin{figure}[H]
    \centering\includegraphics[width=\textwidth]{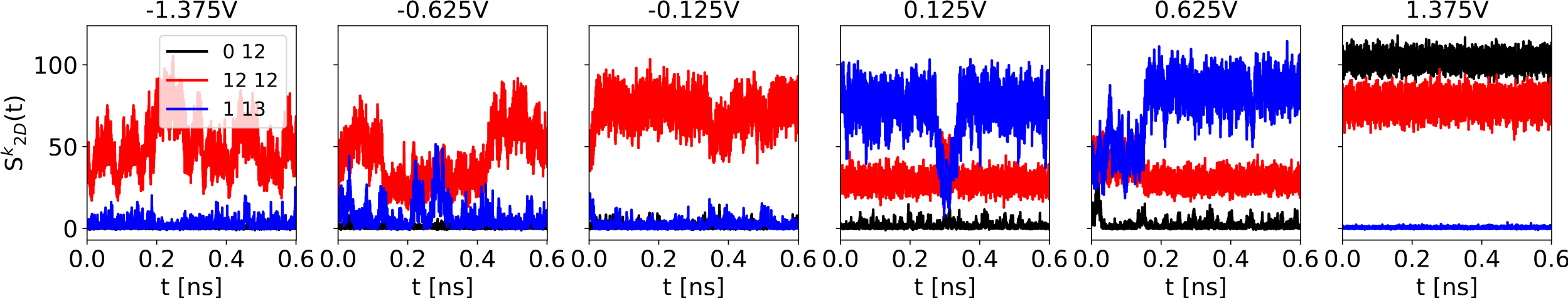}
    \caption{Time evolution of the intensities of three characteristic 2D structure factor peaks:(0,12), (12,12), and (1,13), computed for lithium ions at the interface under three negative (left) and three positive (right) electrode potentials ("medium" system).}
    \label{fig:sk_LiLi_big}
\end{figure}

\begin{figure}[H]
    \centering\includegraphics[width=0.9\textwidth]{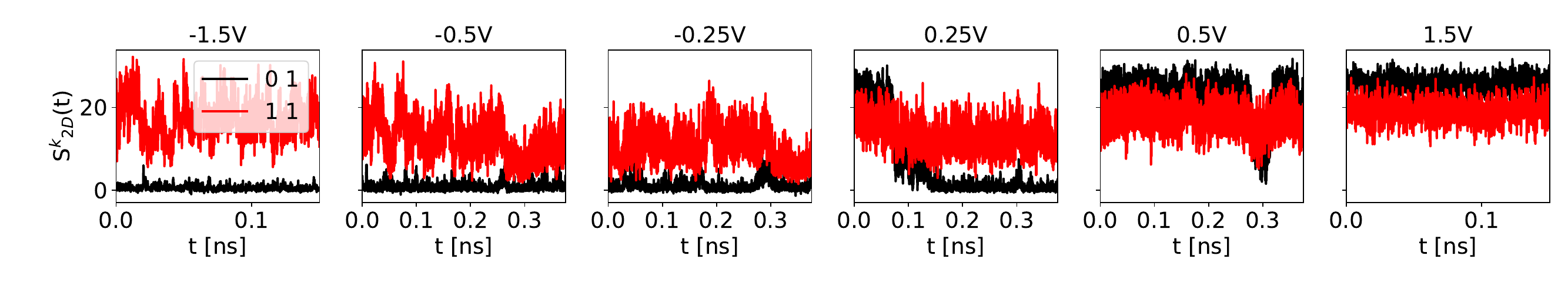}
    \caption{Time evolution of the intensities of two characteristic 2D structure factor peaks:(0,1) and (1,1), computed for lithium ions at the interface under three negative (left) and three positive (right) electrode potentials ("small" system).}
    \label{fig:sk_LiLI_small}
\end{figure}

\begin{figure}[H]
    \centering\includegraphics[width=0.9\textwidth]{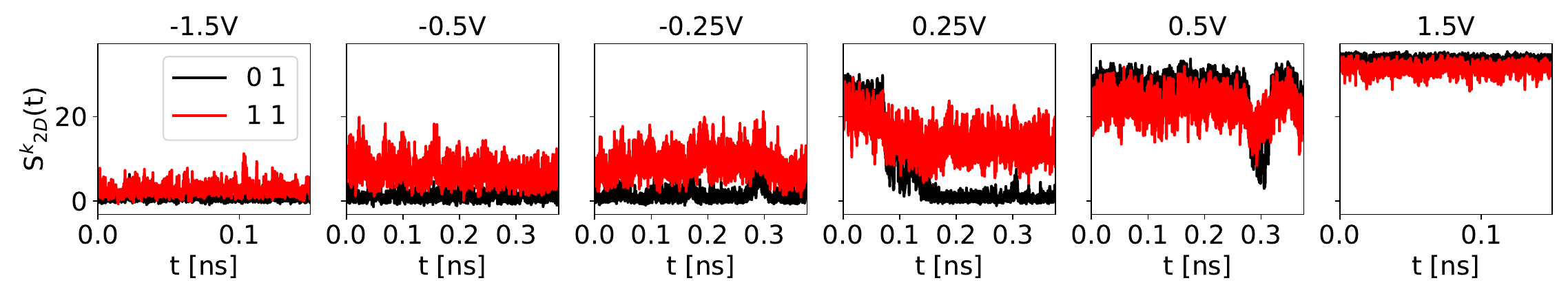}
    \caption{Time evolution of the intensities of two characteristic 2D structure factor peaks:(0,1) and (1,1), computed for chloride ions at the interface under three negative (left) and three positive (right) electrode potentials ("small" system).}
    \label{fig:sk_CLCL_small}
\end{figure}

\begin{figure}[H]
    \centering\includegraphics[width=0.9\textwidth]{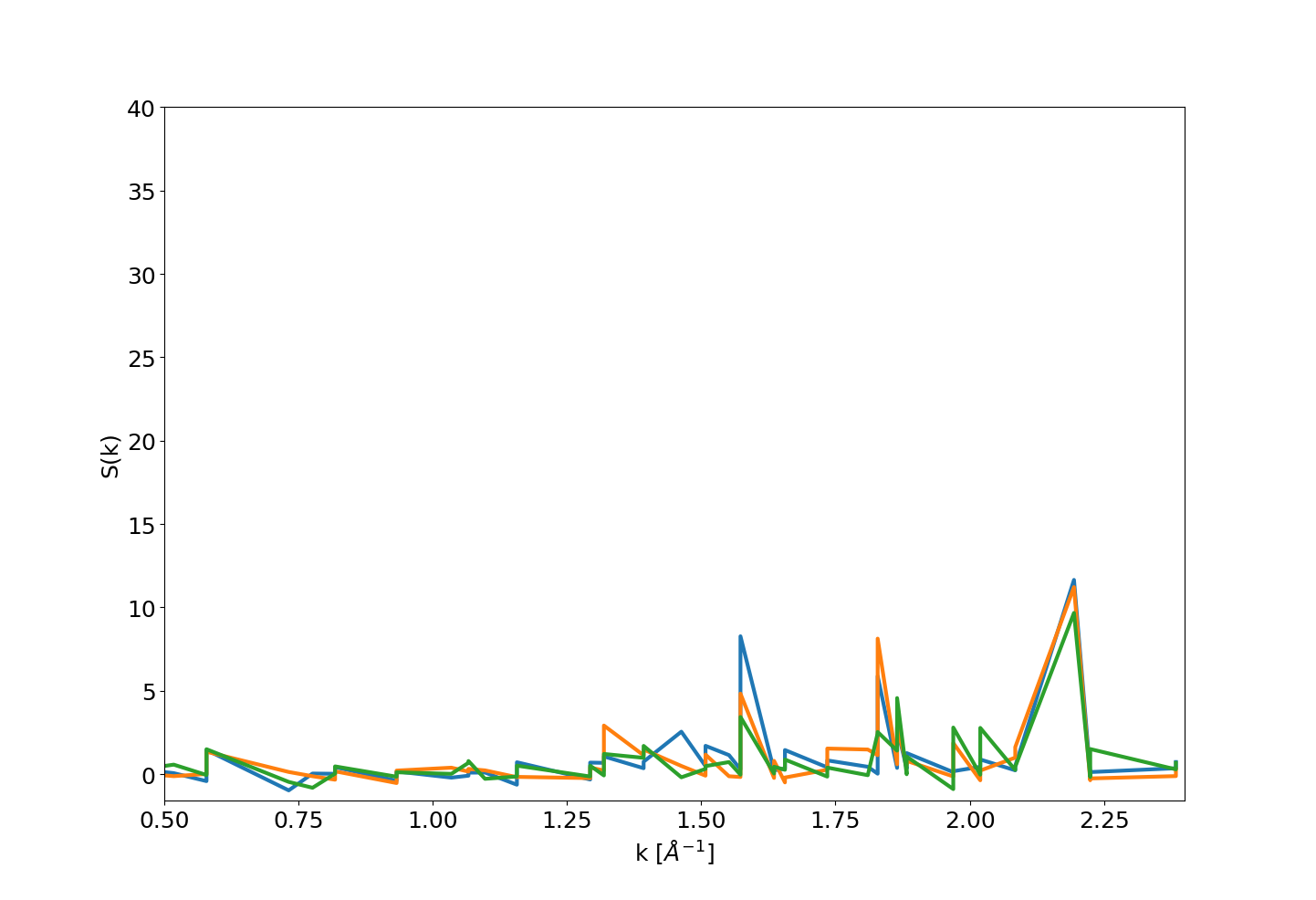}
    \caption{Instantaneous in-plane structure factors for the adsorbed layer for the chloride on the positive electrode at low voltage ("small" system).}
    \label{fig:sk_CLCL_small_total}
\end{figure}

\section{Probability and Free energy}
\begin{figure}[H]
    \centering\includegraphics[width=0.7\textwidth]{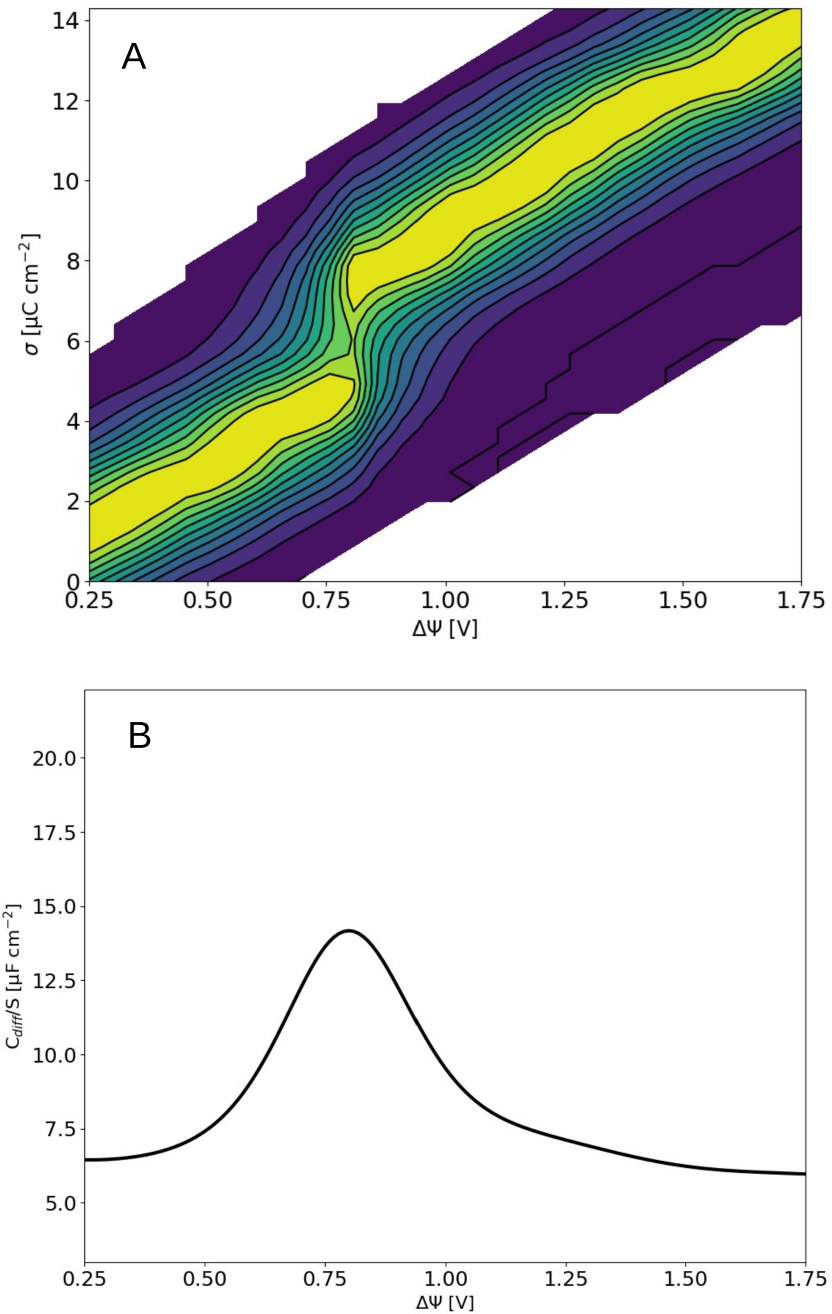}
    \caption{(A) Probability $P(\sigma \mid\Delta\Psi)$ of the charge density on the electrodes at a given applied potential. (B) Differential capacitance as a function of applied voltage. ("small" system)}
    \label{fig:prob_cap_small}
\end{figure}

\end{document}